\documentclass[twocolumn,prl,showpacs,aps,superscriptaddress]{revtex4}

\usepackage{multirow}
\usepackage{graphicx} % for figures
\usepackage{color}

\begin{document}

\date{\today}
\title{How `pairons' are revealed in the electronic specific heat of cuprates}

\author{Yves Noat}

\affiliation{Institut des Nanosciences de Paris, CNRS, UMR 7588 \\
Sorbonne Universit\'{e}, Facult\'{e} des Sciences et Ing\'{e}nierie, 4 place
Jussieu, 75005 Paris, France}

\author{Alain Mauger}

\affiliation{Institut de Min\'{e}ralogie, de Physique des Mat\'{e}riaux et
de Cosmochimie, CNRS, UMR 7590,Sorbonne Universit\'{e}, Facult\'{e} des Sciences et Ing\'{e}nierie, 4 place
Jussieu, 75005 Paris, France}

\author{Minoru Nohara}

\affiliation{Research Institute for Interdisciplinary Science, Okayama University, Okayama 700-8530, Japan}

\author{Hiroshi Eisaki}

\affiliation{Research Institute for Advanced Electronics and
Photonics (RIAEP), National Institute of Advanced Industrial Science
and Technology (AIST), Tsukuba, Ibaraki 305-8568, Japan}

\author{William Sacks}

\affiliation{Institut de Min\'{e}ralogie, de Physique des Mat\'{e}riaux et
de Cosmochimie, CNRS, UMR 7590,Sorbonne Universit\'{e}, Facult\'{e} des Sciences et Ing\'{e}nierie, 4 place
Jussieu, 75005 Paris, France}

\pacs{74.72.h,74.20.Mn,74.20.Fg}

\begin{abstract}

Understanding the thermodynamic properties of high-$T_c$ cuprate
superconductors is a key step to establish a satisfactory theory of
these materials. The electronic specific heat is highly
unconventional, distinctly non-BCS, with remarkable doping-dependent
features extending well beyond $T_c$. The pairon concept, bound
holes in their local antiferromagnetic environment, has successfully
described the tunneling and photoemission spectra. In this article,
we show that the model explains the distinctive features of the
entropy and specific heat throughout the temperature-doping phase
diagram. Their interpretation connects unambiguously the pseudogap,
existing up to $T^*$, to the superconducting state below $T_c$. In
the underdoped case, the specific heat is dominated by pairon
excitations, following Bose statistics, while with increasing
doping, both bosonic excitations and fermionic quasiparticles
coexist.

\end{abstract}

\maketitle

\paragraph{Introduction}

More than thirty years after the discovery of cuprate
superconductivity by Bednorz and M\"{u}ller \cite{ZPhys_Bednorz1986},
the challenge persists to describe their transport, spectroscopic
and thermodynamic properties in a coherent and satisfactory way. In
particular, the thermodynamic properties are of high interest giving
access to the fundamental excitations of the system at equilibrium.

The pioneering studies of the specific heat
\cite{JJAPhys_Kitazawa1987} showed the inherent difficulty to
separate the electronic from the phonon contribution in a variaty of
cuprates near the superconducting transition (see the review of
Fisher et al.\cite{JSC_Fisher1988} and references therein). However,
the innovative work of Loram et al. \cite{PhysicaC_Loram1989,
PhysicaC_Loram1994} showed convincingly that the electronic part of
the specific heat $C_e(T)$ is highly unconventional and deviates
markedly from the BCS behavior \cite{PR_BCS1957}.

\begin{figure*}
\includegraphics[width=\textwidth]{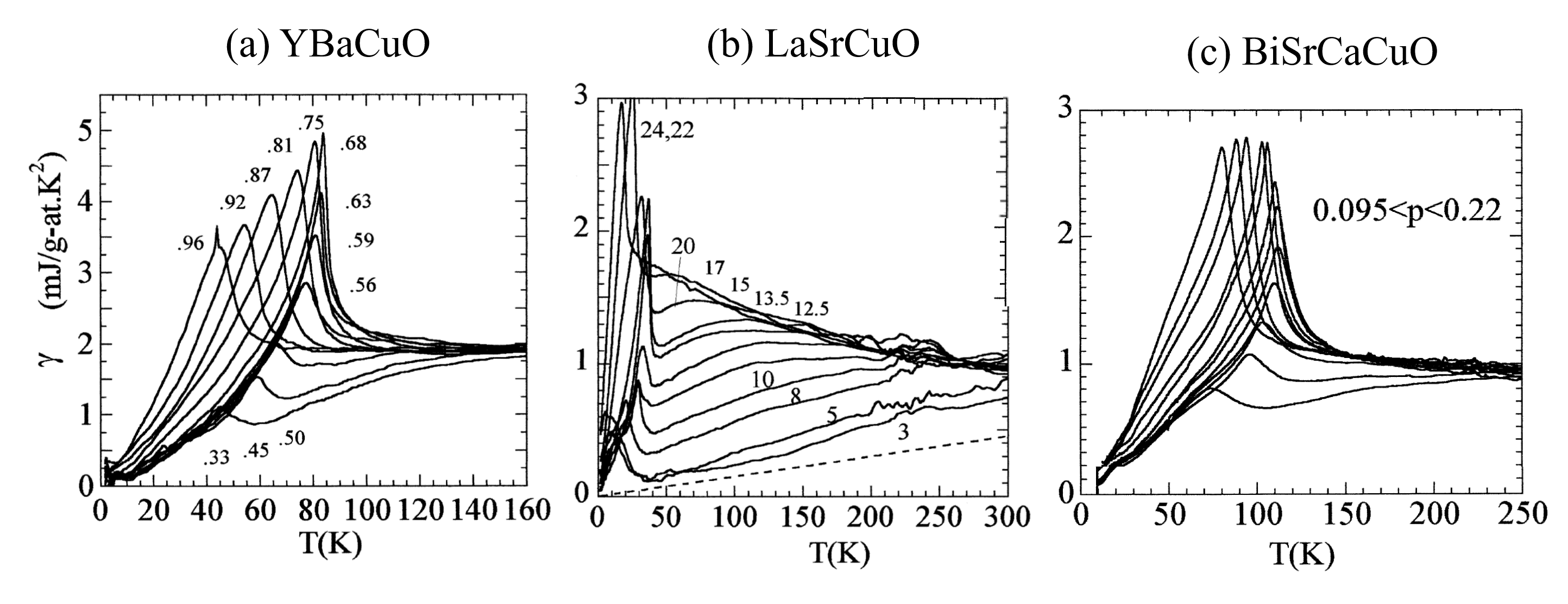}
\caption{(Color online) $\gamma$ coefficient measured as a function
of temperature for three different cuprates,
Y$_{0.8}$Ca$_{0.2}$Ba$_2$Cu$_3$O$_{6+\delta}$ (a),
La$_{2-x}$Sr$_x$CuO$_4$ (b) and (20\% Pb doped, 15\% Y doped)
Bi$_2$Sr$_2$CaCu$_2$O$_{8+\delta}$ (c), by Loram et al.
\cite{JphysChem_Loram2001}.} \label{Fig_Cv data}
\end{figure*}

Below $T_c$, low temperature measurements have demonstrated the
d-wave symmetry of the order parameter
\cite{PhysicaC_Momono1994,PhysicaC_Momono1996,PhysicaC_Mirza1997,
PhysicaC_Nohara2000,PRL_Moler1994,JPSJ_Nohara2000}. The global shape
of $C_e(T)$ is strongly doping dependent and electronic signatures
extend well beyond the critical temperature, especially in the
underdoped case (see Fig.\,\ref{Fig_Cv data}). The pseudogap (PG) in
the normal state, revealed by RMN
\cite{PRL_Alloul1989,PRL_Warren1989} tunneling
\cite{PRL_renner1998_T} and angle-resolved photoemission
spectroscopy (ARPES) \cite{Nat_Ding1996,Sci_Loeser_1996}
experiments, is also evidenced by the specific heat
\cite{JPhysChemSol_Loram1998,JphysChem_Loram2001}. Its relation to
the superconducting state (preformed pairs, coexisting or competing
orders) is still debated. Also, it remains to be clarified whether
the pseudogap is present all along the superconducting (SC) dome or
ends at a quantum critical point for $p\approx 0.2$ above which
(i.e. in the overdoped regime) a Fermi liquid behavior and BCS
superconductivity would be recovered.

A key question is whether these features in the specific heat
can be well understood in the framework of a `preformed pair' model,
wherein $T^*$ is the onset of pair formation, and not a competing
gap as in \cite{JphysChem_Loram2001}. In this article, we use the
pairon model to calculate the electronic specific heat of cuprates,
as a function of temperature ($T$) and hole doping ($p$). The model
allows to describe the main features observed in the specific heat
experiments as a function of $T$ and $p$. We show that the
superconducting transition is governed by pairon excitations
following Bose statistics throughout the phase diagram. Whereas such
excitations dominate at low doping, there is a coexistence of both
pairon and quasiparticle excitations in the overdoped regime.

\paragraph{Pairon model}

We have recently proposed that superconductivity in cuprates can be
explained by the formation of a new quantum object. The pairons are
bound pairs of holes which form in their local antiferromagnetic
environment \cite{EPL_Sacks2017}. They naturally reconcile
antiferromagnetism and Cooper pairing, two normally antagonistic
phenomena. The binding energy $\Delta_p$ is dictated by the spin
exchange interaction $J$ and its doping dependence is linear with
$p$, in agreement with many experiments including ARPES and
tunneling \cite{RepProgPhys_Hufner2008}. Condensation of pairons
arises due to their mutual interaction, giving rise to a collective
quantum state with superconducting properties.

At odds with the BCS case, the critical temperature is not directly
proportional to the gap energy but rather to the pair-pair
interaction $\beta_c \simeq \,2.2 \,k_BT_c$. This interaction
parameter depends on both the density of pairons, which follows the
doping, and on their binding energy $\Delta_p$, the latter having
the characteristic temperature scale $T^*$:
\begin{equation}
\beta_c=C\frac{\left(p-p_{min}\right)}{p_c}\Delta_p
\label{Beta_c}
\end{equation}
where $p_{min}\approx$0.05 is the value at the SC onset and where
$C=0.9$ for  Bi$_2$Sr$_2$CaCu$_2$O$_{8+\delta}$, as deduced from
fits of tunneling data \cite{EPL_Sacks2017}. The critical doping
$p_c \approx 0.27$ is directly related to the pairon size
\cite{EPL_Sacks2017}. While the gap energy is analogous to the
Cooper pairing between two fermions, the quantity $\beta_c$ arises
from an additional four fermion term in the hamiltonian
\cite{SciTech_Sacks2015} which couples different pair states. This
explains why a pairing gap can persist above $T_c$, without SC
coherence, being linked to the higher temperature $T^*$.

In our picture, the increase of the pairon density with doping
opposes the decrease of their binding energy, giving rise to the
dome shape for the critical temperature. Both parameters depend on a
single energy scale, the antiferromagnetic exchange energy $J$
\cite{EPL_Sacks2017} and one length scale $\xi_{AF}$
\cite{EPL_Noat2019}. The pairon model hamiltonian allows to
calculate the spectral function as well as the density of states in
excellent agreement with the experimental tunneling spectra
\cite{EPJB_Sacks2016} as well as ARPES, as a function of temperature
and doping \cite{Jphys_Sacks2018}.

\paragraph{Elementary excitations}

\begin{figure*}
\includegraphics[width=\textwidth]{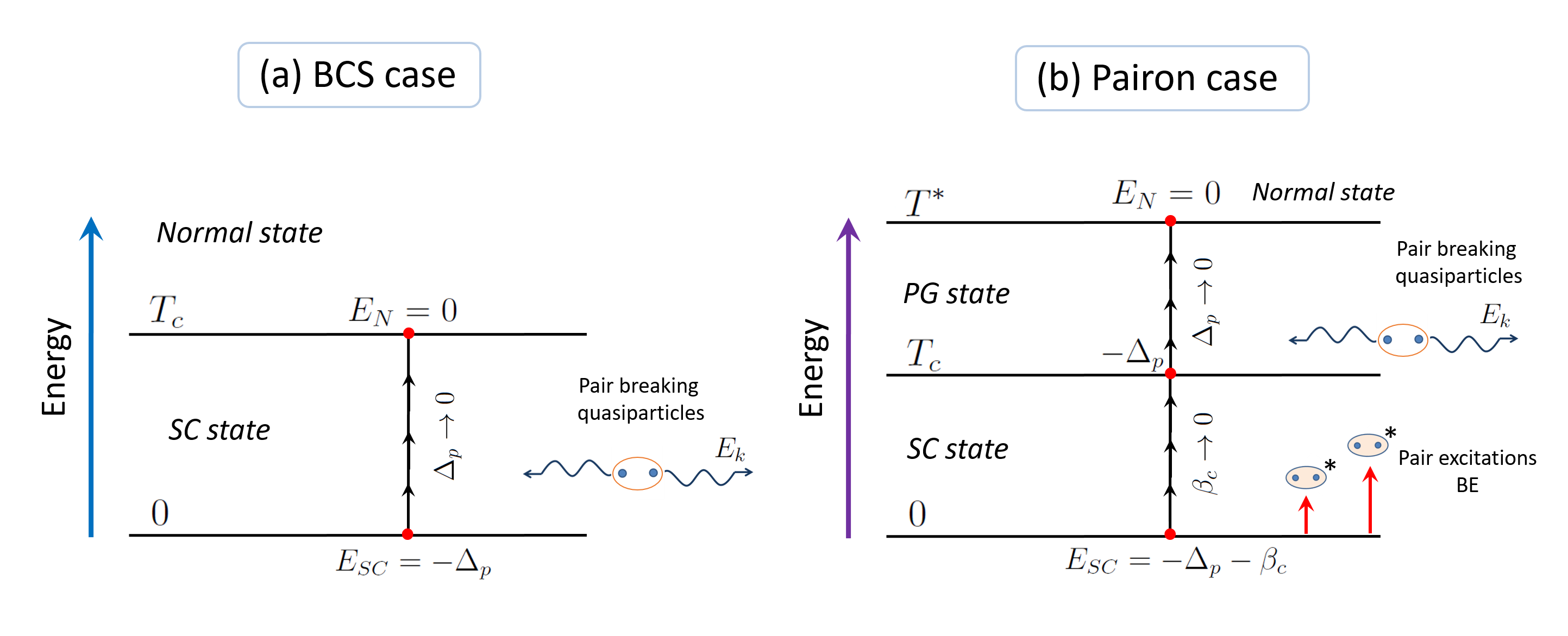}
\caption{(Color online) a) Energy diagram in the BCS case. The
elementary excitations are quasiparticles (fermionic type).  b)
Energy diagram in the pairon model where two kinds of excitations
are present, bosonic (pairon excitation) and fermionic
(quasiparticles) types. While in the BCS case the condensate is
described by a single parameter $\Delta_p$, two energy scales,
$\Delta_p$ and $\beta_c$, are needed to describe the cuprates.}
\label{Fig_Diagram_ener}
\end{figure*}

In conventional superconductors, superconductivity arises due to
electron-electron interaction via phonon exchange resulting in an
energy gap $\Delta_p$ in the excitation spectrum. As shown in the
energy diagram Fig.\,\ref{Fig_Diagram_ener}a, the elementary
excitations from the ground state are quasiparticles of energy
$E_k$, i.e. exotic fermions following Fermi-Dirac statistics. This
results in the familiar quasiparticle density of states with a
temperature dependent gap, the order parameter, which vanishes at
the critical temperature.

In cuprates, as mentioned previously, condensation of pairons is due
to pairon-pairon interaction which leads to a collective quantum
state having long range SC properties. The total ground state energy
(per pair) $E_c$ is given by $E_c=-\Delta_p-\beta_c$, where the
first term is analogous to BCS and the second term arises due to the
mutual interaction between pairons (see the energy diagram
Fig.\,\ref{Fig_Diagram_ener}b). At zero temperature, all pairons
belong to the superconducting ground state with energy $E_c$. As the
temperature increases, pairons are excited out of the condensate
ground state with an occupation number given by Bose-Einstein
statistics.

It is these thermal excitations of pairons that break long range SC
coherence and not quasiparticle excitations. As a result the
condensation energy weakens with temperature and, at the critical
temperature, the effective interaction energy is zero. This is
precisely the pseudogap state where incoherent pairons, with energy
gap $\Delta_p(T_c)$, survive (as indicated in
Fig.\,\ref{Fig_Diagram_ener}b) whereas superconductivity no longer
exists. Further rising temperature implies the familiar pair
breaking into quasiparticles thus leading to the normal state near
$T^*$.

Thus, contrary to BCS where only fermionic excitations are
responsible for the destruction of the SC order, here the bosonic
character of pairons is the key effect. This conclusion was already
discussed in the context of tunneling and ARPES spectra
\cite{EPJB_Sacks2016,Jphys_Sacks2018} and now will be borne out in
the present study of the entropy and the specific heat.

In a Bose picture, the density $n_c$ of condensed pairons is given by
\begin{equation}
n_c(T)=n_0-A\int_{\delta}^{\Delta_0}P_0(\varepsilon_i)f_{BE}(\varepsilon_i)d\varepsilon_i
\label{nc(T)}
\end{equation}
where $A$ is a normalization coefficient, $n_0$ is the density of
pairons at $T=0$ ($\propto p/2$), $P_0(\varepsilon_i)$ is the
density of pairon excited states, and
$f_{BE}(\varepsilon)=1/\left(\exp\left(\frac{\varepsilon-\mu_b}{k_BT}\right)-1
\right)$ is the Bose-Einstein distribution. As in our previous work,
we have choosen a lorentzian form for the density of pairon excited
states:
$P_0(\varepsilon_i)=\sigma_0^2/\left[(\varepsilon_i-\beta_c)^2+\sigma_0^2\right]$,
where $\sigma_0$ is the width of the distribution. Although this
distribution is written differently from our previous work, it is in
fact a change in variable, as detailed below.

The upper limit of integration $\Delta_0$ is the maximum energy of a
pairon while the lower cut-off is a gap in the excitation spectrum
of the pairons ($\delta\sim$2meV) \cite{SciTech_Sacks2015}. Since we
assume a Bose-Einstein like condensation at the critical
temperature, $\mu_b=0$ below $T_c$ while  $\mu_b(T)$ must conserve
the total number of pairons above $T_c$. The constraint that
$n_c(T=T_c)=0$ imposes the value of the normalization coefficient
$A$.

Once the condensate is completely depleted, the total energy of the
system is $\sim \Delta_p(T_c)$, the pseudogap state. In the
underdoped regime, up to the optimal doping, the antinodal gap
varies very little below $T_c$, and consequently the total energy at
the transition is nearly equal to the antinodal gap $\Delta_p(T=0)$.
This clearly illustrates the difference between cuprates and
conventional BCS superconductors: the gap is not the order
parameter.

\paragraph{Entropy calculation}

In a conventional superconductor, the elementary excitations of the
condensed state are quasiparticles arising from the dissociation of
Cooper pairs. Here the increase of the entropy originates from two
fundamental processes, namely pairon excitations following Bose
statistics and the dissociation of pairons into quasiparticles
following Fermi-Dirac statistics. These two fundamental processes
are included in the concise expression of the total entropy $S$:
\begin{equation}
S=\sum_i n_i(\varepsilon_i,T) S_i(\varepsilon_i,T)
\label{Eq_entropy}
\end{equation}
where
$n_i(\varepsilon_i,T)=AP_0(\varepsilon_i)f_{BE}(\varepsilon_i)$ is
the density of excited pairons with energy $\varepsilon_i$ and $S_i$
is the associated entropy term. For every pairon excitation energy
$\varepsilon_i$ there is a set of binding energies $\Delta_i$,
associated with Cooper pairs decaying as quasiparticles of energy
$E_k^i=\sqrt{\epsilon_{k}^2+\Delta_i^2}$. We therefore write:
\begin{equation}
S_i(\varepsilon_i,T)=\sum_{\vec{k}}S(E_k^i,T)
\label{Eq_Si}
\end{equation}
The constitutive equation between the pairon energy and the
associated Cooper pairs, $\varepsilon_i(\Delta_i)$ is needed.
Although phenomenological, we have used with success the equation
$\varepsilon_i=\Delta_i-\Delta_p(T,\theta)$. This equation gives the
excitation energy of pairons with respect to the associated Cooper
pairs of energy $\Delta_i$. $\Delta_p(T,\theta)$ is assumed to be
the average value of the excitation spectrum, at the angle
$\theta$ on the Fermi surface.

\begin{figure}
\includegraphics[width=8.4 cm]{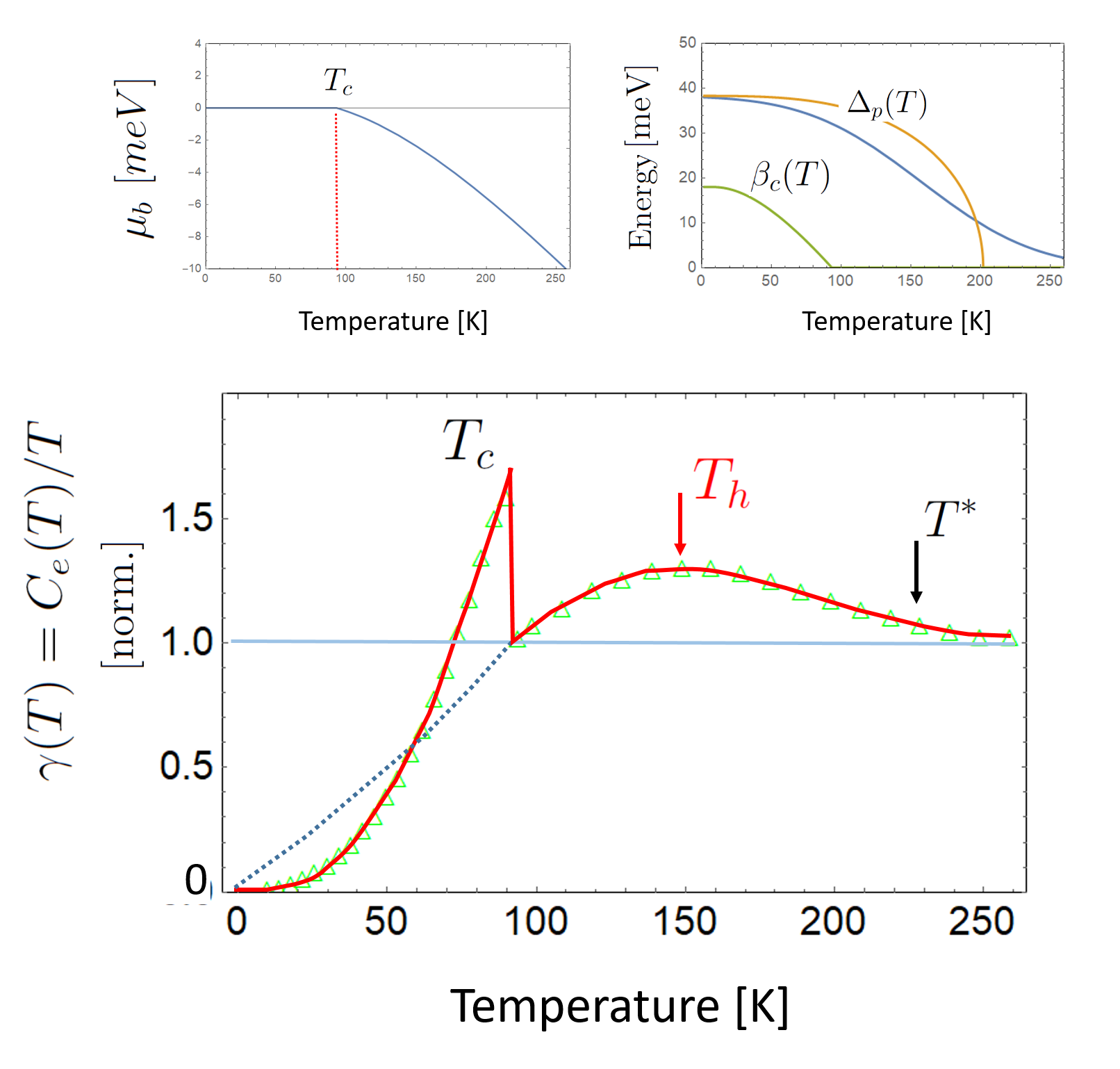}
\caption{(Color online) Lower curve: Electronic specific heat as a
function of temperature at optimum doping, where are indicated the
three characteristic temperatures $T_c$, $T_h$ and $T^*$. Upper
curve, left: chemical potential $\mu_b(T)$ of the pairons. Upper
curve, right: Condensate energy $\beta_c n_c(T)$ which vanishes at
the critical temperature (red curve); Dependence of the gap function
in the antinodal direction used in our calculations (blue curve) as
a function of temperature compared to a gap function following a
BCS-type dependence (orange curve).} \label{Fig_Cv optimum}
\end{figure}

In our previous work, we chose
$\Delta_p(T,\theta)=\Delta_p(T)\cos(2\theta)$ (except for some
angular corrections, due to the spatial extension of the pairons
\cite{EPL_Noat2019}), mainly present in the underdoped regime), with
$\Delta_p(T)$ given by the BCS formula (brown curve, upper right of
Fig.\,\ref{Fig_Cv optimum}. However, such an expression leads to a
discontinuity of the specific heat at $T^*$ which is absent in the
experiments. A precise fit to the data is obtained using a smooth
function which softens the variation of $\Delta_p(T)$, as
illustrated in Fig.\,\ref{Fig_Cv optimum} (upper right, blue curve).

In Eq.\,\ref{Eq_Si}, $S(E_k^i,T)$ is the usual entropy expression
for fermions with energy excitation spectrum $E_k^i$.
\begin{eqnarray}
&S(E_k^i,T)=&-2k_B [f(E_k^i)\ln\left(f(E_k^i)\right)\null \nonumber\\
&&+\left(1-f(E_k^i)\right)\ln\left(1-f(E_k^i)\right)] \
\label{Eq_Si_int}
\end{eqnarray}
with $f(E)$ the Fermi-Dirac function. Summing over all $\vec{k}$
values, gives
\begin{equation}
S_i(\varepsilon_i,T)=-k_B N_n \int_{0}^{\infty}d\epsilon_k\int_{0}^{2\pi}d\theta \,S(E_k^i,T)\ ,
\label{Eq_Si_sumk}
\end{equation}
where $N_n$ is the normal density of states at the Fermi level.
Finally, we replace the discrete sum over the states by an integral
$\sum_i\approx \int d\varepsilon_i P_0(\varepsilon_i)$, and get
\begin{equation}
S=\int_{\delta}^{\Delta_0} d\varepsilon_i \,P_0(\varepsilon_i)\, n_i(\varepsilon_i,T)\, S_i(\varepsilon_i,T),
\label{S_intP0}
\end{equation}
 with $P_0(\varepsilon)$ being the density of pairon excited states.

\begin{figure}
\includegraphics[width=8.4 cm]{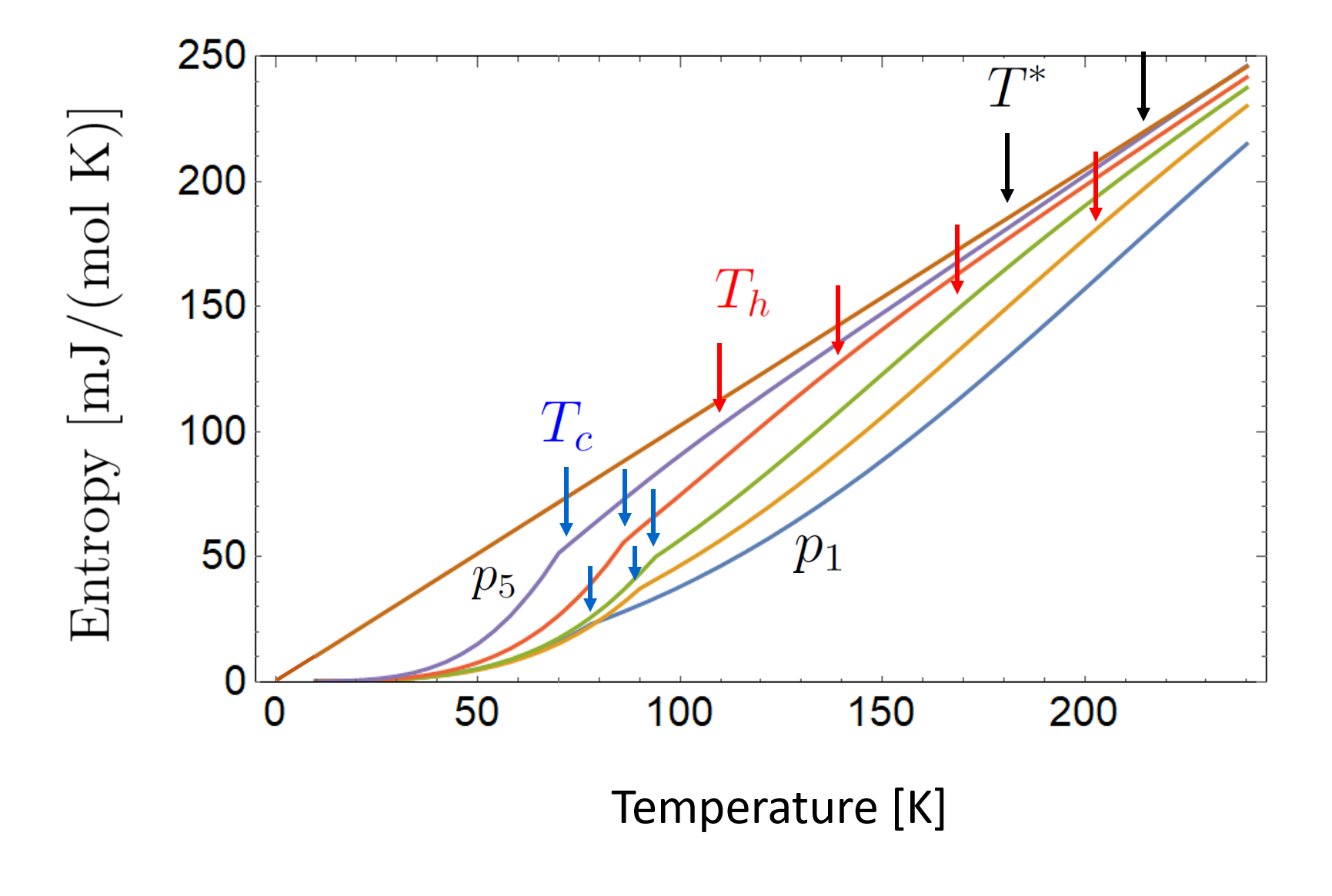}
\caption{(Color online) Entropy as a function of temperature for
different doping values ($p_1=0.11$, $p_2=0.135$, $p_3=0.16$,
$p_4=0.185$, $p_5=0.21$). Red, blue, and black arrows respectively
indicate the position of the critical temperature $T_c$, the hump
temperature $T_h$, and the pseudogap temperature $T^*$.}
\label{Fig_Entropy}
\end{figure}

\paragraph{Results for the entropy and the specific heat}

The entropy as a function of temperature for different doping values
is plotted in Fig. \ref{Fig_Entropy}. At high temperature ($T\gg
T^*$), the entropy $S(T)$ is linear as a function of temperature,
which is characteristic of metallic behavior. This result can be
seen as a limit of Eq.\,\ref{Eq_Si} where the gap energy vanishes
and $S_i$ equals the free electron entropy, and using particle
conservation $\sum_i n_i(\varepsilon_i,T)=n_0$. At lower temperature
($T<T^*$), $S(T)$ departs from the linear asymptote and stays below
the metallic value. This is a direct consequence of the formation of
pairons which, as a result of increasing order, lowers the entropy
relative to the metallic case.

Furthermore, a smooth but marked change of the $S(T)$ curvature is
observed at the temperature $T_h$, located between $T^*$ and $T_c$.
It corresponds to a maximum of quasiparticle excitations which
occurs at the {\it inflection} point of the gap function
$\Delta_p(T)$ (see Fig.\,\ref{Fig_Cv optimum}, right upper curve).
At even lower temperature, the progressive condensation of pairons
is manifested by the lowering of the entropy with an abrupt change
in the slope of $S(T)$ at $T_c$. Below $T_c$, the entropy decreases
rapidly to zero and vanishes at absolute zero.
% The corresponding peak in $C_e(T)$ corresponds to a maximum of quasiparticle excitations which occurs at the inflexion point of the curve $\Delta(T)$, Fig.\,\ref{Fig_Cv optimum}.

The specific heat is obtained directly from the entropy according to
the relation $C_e(T)=T\frac{dS}{dT}$. The standard quantity $\gamma
(T)=C_e(T)/T$ is plotted in Fig.\,\ref{Fig_Cv optimum}, lower curve,
for optimum doping. At low temperature, $\gamma(T)$ is very small
and then increases rapidly due to pairon excitations and reaches a
maximum at the transition $T_c$, where all pairons are excited out
of the condensate, $n_c(T_c)=0$, as illustrated in Fig.\,\ref{Fig_Cv
optimum}, right upper curve. The sharp peak in $\gamma (T)$,
followed by the discontinuity, is therefore associated with the
disappearance of the condensate.

At higher temperature, a smooth hump is observed at $T_h$
corresponding to a maximum of quasiparticle excitations, seen as the
point of inflection in the entropy. This effect has been discovered
and studied by Matzuzaki et al. \cite{JphysJapSol_Matzusaki2004} and
is discussed below. Finally, a constant is recovered for $T\gg T^*$
corresponding to the metallic behavior. Note that the return to the
normal state, corresponding to the vanishing of the pseudogap state,
is smooth and continuous.

Returning to the question of the SC state, the dashed curve in
Fig.\,\ref{Fig_Cv optimum} is obtained for incoherent pairons with
no superconducting transition. Examining $\gamma(T)$ one sees
clearly the conservation of the relative area compared to the
background. While BCS superconductivity emerges from the normal
state (wherein $\gamma(T)=\gamma_N$, $\gamma_N$ being the gamma
coefficient in the normal state), here it emerges from an incoherent
PG state of preformed pairons. We stress that this incoherent state
is responsible for the unconventional background of $S(T)$ and
$C_e(T)$ observed in the experiments.

%One noted that the chemical potential is zero below $T_c$ and non-zero above as a result of conservation of the number of pairons.
\begin{figure}
\includegraphics[width=8.4 cm]{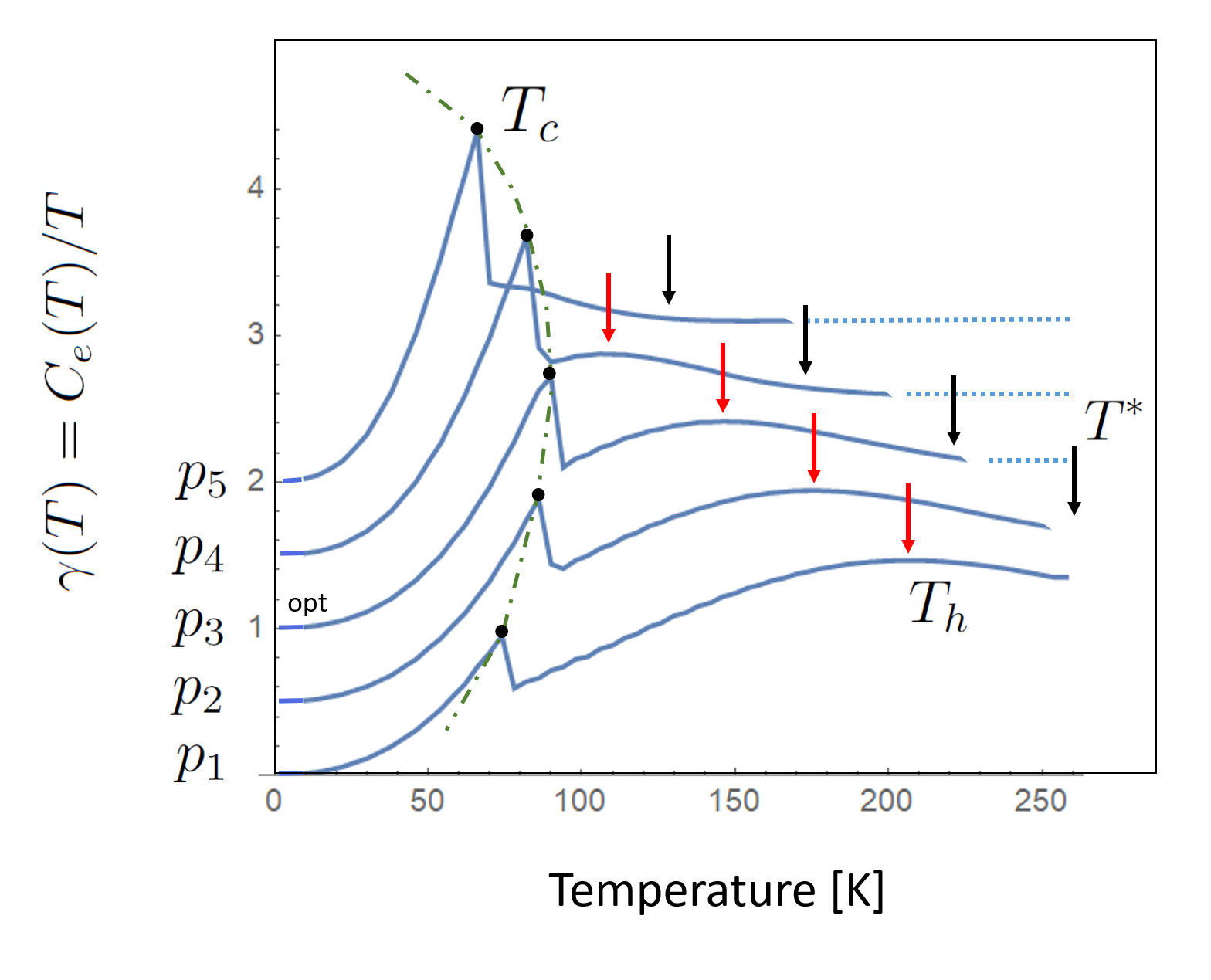}
\caption{(Color online) Lower curve: Electronic specific heat as a
function of temperature for different doping values all over the
phase diagram ($p_1=0.11$, $p_2=0.135$, $p_3=0.16$, $p_4=0.185$,
$p_5=0.21$). $T_c$, $T_h$ and $T^*$, are respectively indicated by
the dashed curve and red and black arrows. The dotted horizontal
line is the asymptotic limit $\gamma(T)=\gamma_N$ reached in the
normal state. The curves have been shifted vertically for clarity.}
\label{Fig_Cv doping}
\end{figure}

\begin{figure}
\includegraphics[width=8.4 cm]{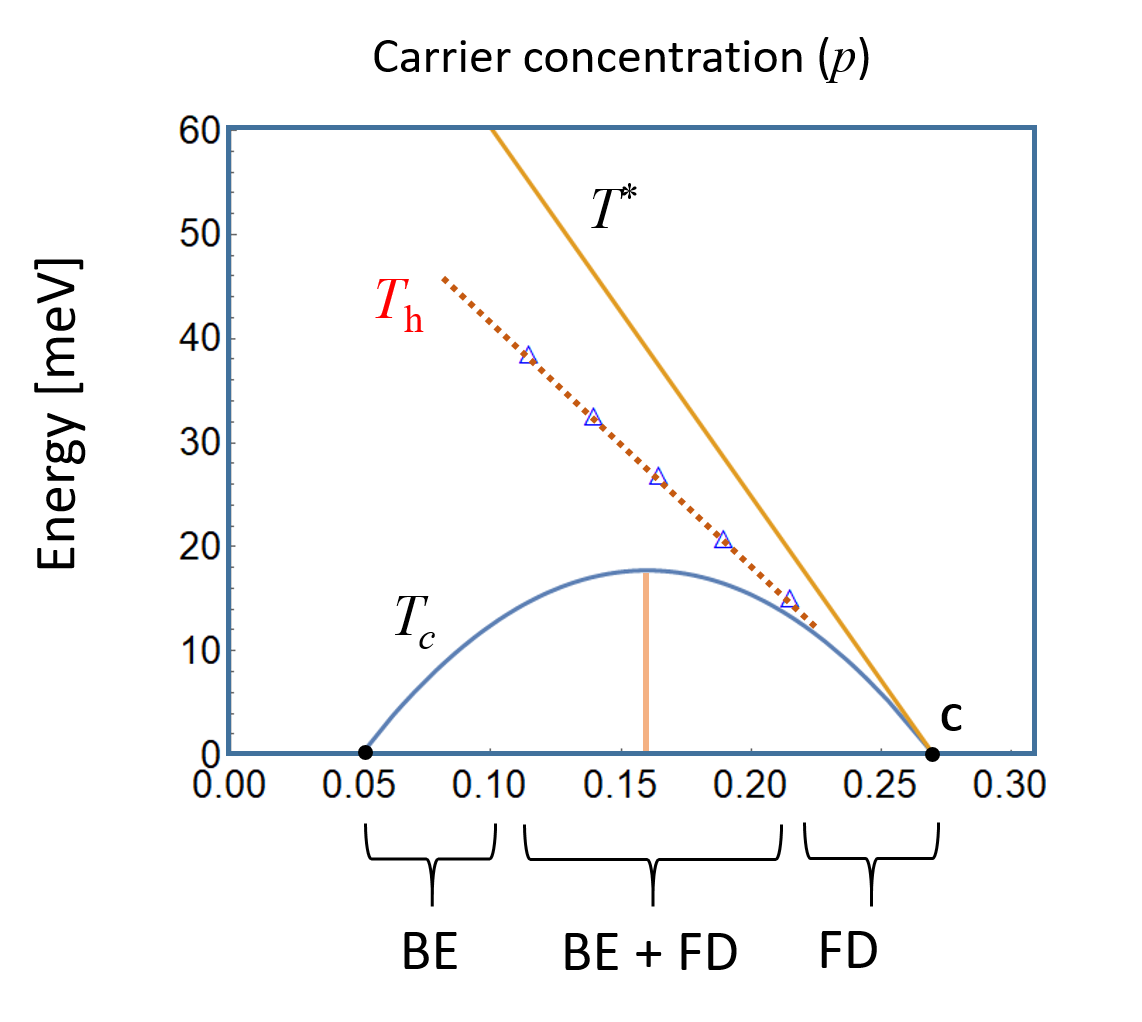}
\caption{(Color online) Phase diagram showing the superconducting
dome $T_c$ ($\propto \beta_c$), the pseudogap temperature $T^*$
($\propto \Delta_p$) as a function of hole doping $p$. The
specific heat `hump' temperature $T_h$, scaling perfectly with
$T^*$, is also plotted. The type of excitations which dominate in
the different doping regime is indicated below. Bose-Einstein (BE)
excitations in the underdoped regime, Bose-Einstein and
quasiparticle fermionic excitations (BE+FD) around the optimum
doping, and fermionic type excitations (FD) in the highly overdoped
region.}\label{Fig_Phasediagram}
\end{figure}

We have also calculated $C_e(T)$ for various doping values along the
phase diagram, as summarized in Fig.\,\ref{Fig_Cv doping}. In
agreement with experiments we find that the shape of the curves is
quite different from underdoped to overdoped regimes. While the
parameters underlying each curve vary monotonically, there is no
simple homothetic relation between the curves. It is evident that
$T_c$ follows a dome shape in agreement with the phase diagram, see
Fig.\,\ref{Fig_Phasediagram}. With increased doping, $T_c$ and $T^*$
approach each other as deduced from ARPES or tunneling spectroscopy
experiments \cite{RepProgPhys_Hufner2008}. Finally, they merge at
the maximum doping $p_c=$0.27, the critical point in our phase
diagram.

The third characteristic temperature $T_h$
\cite{JphysJapSol_Matzusaki2004} is a prominent feature of the
present calculation. As seen in Fig.\,\ref{Fig_Cv optimum}, for
optimum doping, the hump is roughly midway between $T^*$ and $T_c$.
At this doping, the quasiparticle formation is quite distinct from
the SC transition at $T_c$. However for larger $p$, the hump moves
down towards $T_c$ and merges progressively with the `bosonic'
discontinuity at $T_c$. In the overdoped regime, $T_h$ is no longer
visible. This behavior is particularly evident in experimental
results on La$_{2-x}$Sr$_x$CuO$_4$ (Fig.\,\ref{Fig_Cv data}) and
also in (20\% Pb and 15\% Y doped)
Bi$_2$Sr$_2$CaCu$_2$O$_{8+\delta}$, although the effect is less
pronounced.

\paragraph{Discussion}

It is remarkable that the entropy and the electronic specific heat
can be described by the same set of equations (Eq. \ref{Eq_entropy},
\ref{Eq_Si} and \ref{Eq_Si_int}), regardless of the carrier
concentration all along the phase diagram. Although the detailed
shape of $C_e(T)$ varies significantly as a function of doping,
quite surprisingly, it can be explained by the same underlying
mechanisms. Moreover, the conclusions are in agreement with our
previous work, where we deduced a universal phase diagram involving
the most relevant parameters, the pairon binding energy and their
mutual interactions \cite{EPL_Sacks2017}. The fundamental length
scale $\xi_{AF}$ and energy scale $J$, the antiferromagnetic
exchange energy, are at the heart of the wide range of phenomena as
seen in tunneling and ARPES. Such a simplified picture is now in
qualitative agreement with the specific heat.

In our view, there is no abrupt transition as a function of doping
from underdoped to overdoped sides, but a continuous evolution. The
apparent change of behavior seen in the specific heat, just as in
tunneling and ARPES experiments, reflects the dual nature of the
excitations of the system, fermionic and bosonic type. Above $T_c$,
the characteristic temperature $T_h$ corresponds to the point where
the change of population of quasiparticle excitations reaches a
maximum, which occurs at the inflection point of $\Delta_p(T)$.
Above this temperature, as illustrated in Fig.\,\ref{Fig_Cv
optimum}, $\Delta_p(T)$ is rapidly decreasing.

It is close to the node that quasiparticle excitations from decaying
pairons dominate. This is the Fermi-arc contribution included in our
calculation, being directly proportional to $T_c/T^*$, as we have
shown in \cite{Jphys_Sacks2018}. In the underdoped regime, for
$T<T_c\ll T_h$ pairon excitations dominate in entropy and the
specific heat (Fig.\,\ref{Fig_Entropy} and Fig.\,\ref{Fig_Cv
doping}). However, the composite nature of pairons is key. As the
doping increases, $T_h$ decreases, then more quasiparticle
excitations are present at lower temperature. Finally, on the higher
doping range, ($T_h\sim T_c$), there is a coexistence of both types
of excitations.

The hump in the specific heat curve is also associated with the
number of the pairon energy states. On the lower-temperature side of
the hump, the specific heat increases because of a large increase of
entropy, according to Eq.\ref{Eq_Si} as more pairon excited states
become populated. Interestingly, this effect is similar to a
Schottky anomaly. However, in a standard Schottky anomaly, the
decrease of the specific heat on the right side of the hump is the
result of the high population of the excited energy levels. In our
case, it is due to quasiparticle decay of pairons. The hump in the
specific heat can thus be considered as an unconventional Schottky
anomaly.

The present study shows that the temperature dependence of the
thermodynamic quantities can be well described by assuming an
average gap $\Delta_p(T)$, describing the pairon excited states,
which decreases very smoothly and vanishes at the typical
temperature $T^*$ (Fig.\,\ref{Fig_Cv optimum}). This mean field gap
is not due to an hypothetical competing order and shows no sign of
discontinuity neither is it due to SC fluctuations, the energy scale
being too large. Rather, the gap originates from excited pairs above
$T_c$, in agreement with the conclusion raised by Wen et al.
\cite{PRL_Wen2009} from specific heat measuremements in
Bi$_2$Sr$_{2-x}$La$_x$CuO$_{6+\delta}$. Therefore in our model,
there is no phase transition associated with the characteristic
temperature $T^*$.

However, the shape of $\Delta_p(T)$ does imply that a residual
pairon density persists above $T^*$. This residual density is likely
to be too small to have any significant effect on the tunneling and
ARPES spectra. Nevertheless, it remains an interesting and open
question as to their possible detection above $T^*$.

In their stimulating work, Curty et al.\cite{PRL_Curty2002} also
address the calculation of the specific heat and the interpretation
of the phase diagram. They consider an attractive Hubbard-like
hamiltonian with a d-wave local pairing on adjacent sites and then
determine thermodynamic properties in a Ginzburg-Landau/Monte Carlo
approach. Their work reveals that the superconducting state emerges
from  an incoherent phase of pairs. The present work is in
qualitative agreement with a number of their conclusions. Based on
the idea of incoherent preformed pairs, we see that two energy
scales, without a discontinuity of behavior for a wide range of
doping, fits the specific heat and entropy. Curty et al. also stress
the absence of a quantum critical point under the dome. This is also
the case in our work, as illustrated in
Fig.\,\ref{Fig_Phasediagram}, since the pseudogap persists for all
$p$ values, even in the overdoped regime up to the maximum value
$p_c$.

%\begin{multicols}{2}
\paragraph{Conclusion}

We have shown that the thermodynamic properties of cuprates can be
very well described by the formation of pairons, bound pairs of
holes in their antiferromagnetic environment. Superconductivity
emerges from an incoherent state of pairons, the pseudogap state, as
a result of their mutual interactions. Two fundamental temperature
scales can be clearly identified in specific heat experiments. They
correspond to the mean binding energy of pairons $\Delta_p$, related
to the pseudogap temperature $T^*$, and the interaction energy
$\beta_c$, which directly proportional to the superconducting
critical temperature. The peak at the critical temperature in the
specific heat is explained in terms of pairon excitations following
Bose-Einstein statistics which deplete the condensate. In the
underdoped regime, pairon excitations qualitatively dominate in the
specific heat up to the critical temperature. As the doping
increases, the contribution of quasiparticle fermionic excitations
becomes more and more pronounced even below $T_c$: the Fermi-arc
phenomenon. Therefore, both types of excitations coexist,
particularly in the overdoped regime, a unique feature of cuprate
superconductivity.

\vskip 2 mm

\paragraph{Acknowledgments}

One of us (WS) is grateful for visiting researcher and
professorship, at AIST Tsukuba and Okayama University respectively,
during the first half of 2020. Support at AIST was provided by a
JSPS FY2019 {\it Invitational Fellowship for Research in Japan}, and
we acknowledge Professors Hideo Aoki, Shin-ichi Uchida and Kohji
Kishio for very fruitful discussions.

\end{document}